\documentstyle[11pt,paspconf]{article}

\def\etal{{\it et~al.\,}}
\def\nuc#1#2{$^{#1}{\rm #2}$}
\def\sles{\lower2pt\hbox{$\buildrel {\scriptstyle <}
   \over {\scriptstyle\sim}$}}
\def\sgreat{\lower2pt\hbox{$\buildrel {\scriptstyle >}
   \over {\scriptstyle\sim}$}}

\begin{document}

\title{New Insights into Core--Collapse Supernova Theory}
\author{Adam Burrows}
\affil{Department of Astronomy, University of Arizona, Tucson, AZ 85721, USA}

\begin{abstract}

Though it is agreed that the post--bounce, pre--explosion cores of massive
stars are unstable to transonic, Rayleigh--Taylor--like instabilities, the role
convective motions may play in igniting the supernova explosion itself is not yet clear.
Whatever that role, the morphology and nucleosynthetic yields of the explosion, the recoil of the
protoneutron star, the spatial distribution of fresh $^{56}$Ni, and the gravitational
wave and neutrino signatures of the event will be affected in interesting ways.  
I review a subset of the issues raised by the new paradigm and some of the technical
obstacles that remain.  I focus on neutrino transfer, the role of progenitor structure,
kick mechanisms, and the location and potential consequence of various hydrodynamic
instabilities.  

\end{abstract}

\keywords{supernova mechanism, neutrinos, gamma--rays, pulsar proper motions, multi-dimensional hydrodynamics, SN1987A}

\section{Introduction}

Neutrino--driven Rayleigh--Taylor instabilities between
the stalled shock wave and the neutrinospheres are generic feature of core--collapse
supernovae (Bethe 1990; Herant, Benz, \& Colgate 1992; Herant \etal\ 1994; Burrows, Hayes, \& Fryxell 1995 (BHF);
Janka \& M\"uller 1996; Mezzacappa \etal\ 1996).  
Supernovae must explode
{\it aspherically}, and this broken symmetry is imprinted upon the ejecta and character of the blast,
as well as its signatures.   Some consequences of asphericity include significant gravitational 
radiation (Burrows \& Hayes 1996; M\"onchmeyer \etal\ 1991),
natal kicks to nascent neutron stars (Burrows \& Hayes 1996; Woosley 1987), induced rotation
(BHF), mixing of iron--peak and r--process nucleosynthetic products,
the generation and/or rearrangement of pulsar magnetic fields,
and, in extreme cases, jetting of the debris (see D. Wooden, R. Stathakis, this volume).  

There is no consensus yet on the centrality of ``convection'' to the
mechanism of the explosion itself, with some deeming it either pivotal (Bethe 1990; Herant, Benz, \& Colgate 1992), 
potentially important (BHF; Janka \& M\"uller 1996; Mayle \& Wilson 1988),
or unimportant (Mezzacappa \etal\ 1996).  Nevertheless, {\it all} agree on the existence
of convection in the gain region of the stalled protoneutron star.
A gain region is a prerequisite for the neutrino--driven mechanism (Bethe \& Wilson 1985). For heating to exceed 
cooling in steady--state accretion, the 
entropy gradient must be negative, and, hence, unstable.  Therefore, a gain region is always convective.
In order to achieve quantitative agreement with the variety of observational constraints (explosion energy,
residual neutron star masses, $^{56,57}$Ni and ``$N = 50$'' peak yields, 
halo star element ratios,
neutron star proper motions, etc.), the ``final'' calculations 
must be done multi--dimensionally.
While if it can be shown that 1--D spherical models do explode after some delay, the true duration
of that delay, the amount of fallback, and the energetics of the subsequent explosion must be influenced by the overturning
motions that can not be captured in 1--D.  Convection changes not only the character of the hydrodynamics, but the entropies 
in the gain region and the ``efficiency'' (Herant, Benz, \& Colgate 1992) of neutrino
energy deposition that is the ultimate driver of the explosion (BHF; Janka 1993; Bethe \& Wilson 1985).
Furthermore, implicit in a focus
on 1--D calculations is the assumption that multi--D effects could only help, that they do not
thwart explosion.  
Hence, the belief that spherically--symmetric calculations are germane depends upon insights newly obtained
from the multi--dimensional simulations.  Nevertheless, it will be an important theoretical exercise to ascertain whether 
1--D models with the best physics and numerics can explode, if only because such has been a goal 
for decades.  The ``viablility'' of 1--D models will be influenced in part by the transport algorithm employed
(multi--group, flux--limited, full transport, diffusion), the microphysics (opacities and source terms
at high and low densities), the effects of general relativity, the equation of state, 
convection in the inner core that can boost the driving neutrino luminosities (Burrows 1987; Keil, Janka, \& M\"uller 1996;
Mayle \& Wilson 1988), and the inner density structure
of the stellar progenitor.

\section{The Role of Progenitor Structure in Core--Collapse Supernovae}

The density profiles of collapsing cores are functions of progenitor ZAMS mass (Weaver \& Woosley 1995; Nomoto \& Hashimoto 1988).  
The less massive progenitors (8 M$_\odot$ \sles\, $M$ \sles 15--20(?)M$_\odot$)  
have compact cores with tenuous envelopes, while the more
massive progenitors (20 (?) M$_\odot$ \sles\, $M$) have massive cores and dense envelopes.
Figure 1 depicts the density profiles for some of the progenitor models in the literature.

\begin{figure}
\vspace{3.5in}
 
\caption{Density (in gm cm$^{-3}$) versus interior mass (in solar masses) for some of the massive progenitor models of 
Weaver \& Woosley (1995) (s\#s) and Nomoto \& Hashimoto (1988) (helium6.0, helium3.0). The edges of 
the iron, silicon, oxygen, and helium zones are indicated by the
symbols. The more massive models have higher densities at large interior masses exterior to 2.5 M$_{\odot}$.}
 
\label{fig-1}
\hbox to\hsize{\hfill\includegraphics{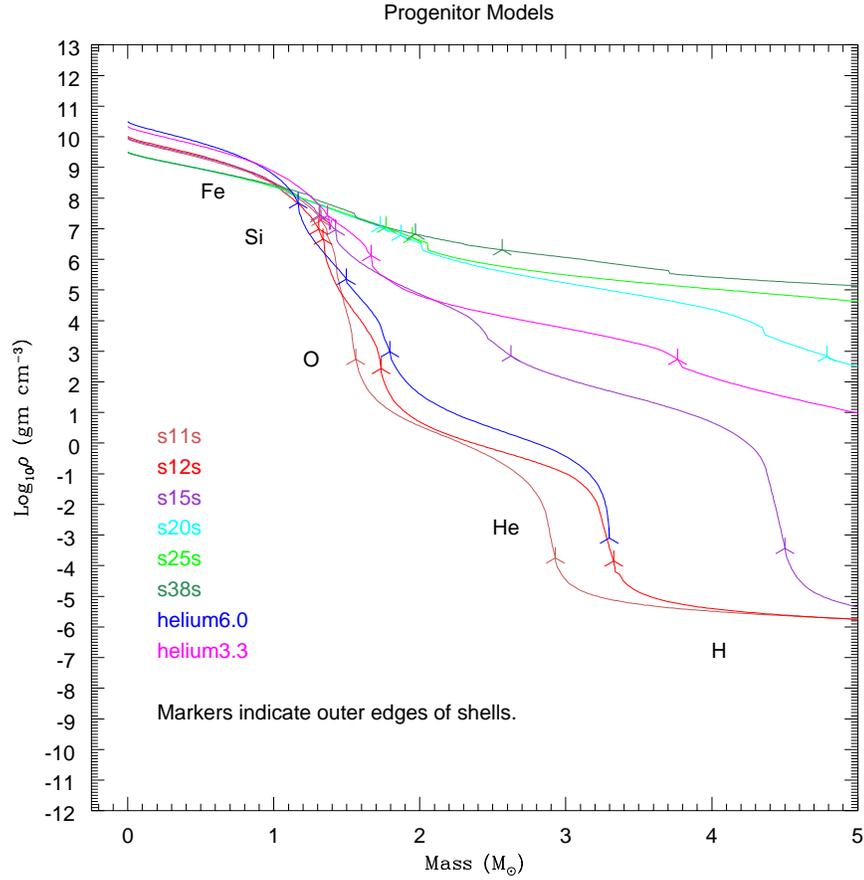}\kern+0in\hfill}
 
\end{figure}

In this model set, the density at 2 M$_\odot$ interior mass for the 38 M$_\odot$ progenitor is {\it six} orders
of magnitude greater than that at the same point for the 11 M$_\odot$ progenitor.
The structure of the core determines the mass accretion rate after bounce, 
as well as the binding energy of the inner mantle that an explosion must overcome to leave a neutron
star rather than a black hole (see Figure 2; BHF; Timmes, Woosley, \& Weaver 1996). 
A high mass accretion rate may smother, inhibit, or delay explosion.
In a sense, a progenitor's inner density structure determines the outcome of its core's collapse.
The cut between compact and extended cores may be at
a different ZAMS mass (Bazan \& Arnett 1994), but a bifurcation into two classes seems to be an 
important ingredient of the collapse story.
In the calculations of Burrows, Hayes, \& Fryxell (1995), while the 15 M$_\odot$ progenitor exploded,
the 20 M$_\odot$ progenitor did not.   However, the explosion in that calculation, 
as well as in that of Herant {\em et al.} (1994),  occurred too
early to be consistent with nucleosynthetic constraints, ejecting ten (BHF) and one hundred (Herant \etal\ 1994) 
times as much neutron--rich
material as the data allow.  In and of itself, better transport will no doubt positively influence the duration of the delay to
explosion (Mezzacappa \etal\ 1996; Burrows \& Pinto 1997) and the resulting 
nucleosynthetic yields, but more attention needs to be paid to progenitor stellar evolution.
In this regard, Bazan \& Arnett (1994) are studying oxygen and silicon burning prior
to collapse in 2--D.   In the oxygen--burning zone, they are obtaining Mach \#'s 
near 25\% and anisotropies in the composition, density, and velocity
that are far in excess of those inferred using the mixing--length prescription.  
These calculations suggest that the final word has not been uttered concerning supernova progenitor structures.

\begin{figure}
\vspace{5.5in}
\caption{This figure depicts the exterior binding energies of some of the Weaver \& Woosley (1995)
progenitor structures versus interior mass.  A larger ZAMS mass object has a larger binding energy at 
a given mass point. Short names for the Weaver \& Woosley (1995) models are shown in a column in the right--hand--corner
of the figure, the numbers in the names representing the mass of the model in solar masses.  
According to this figure, the most massive models
should be difficult to explode, without leaving behind protoneutron star masses that are in excess of the
maximun stable neutron star mass.
}

\label{fig-2}
\hbox to\hsize{\hfill\includegraphics{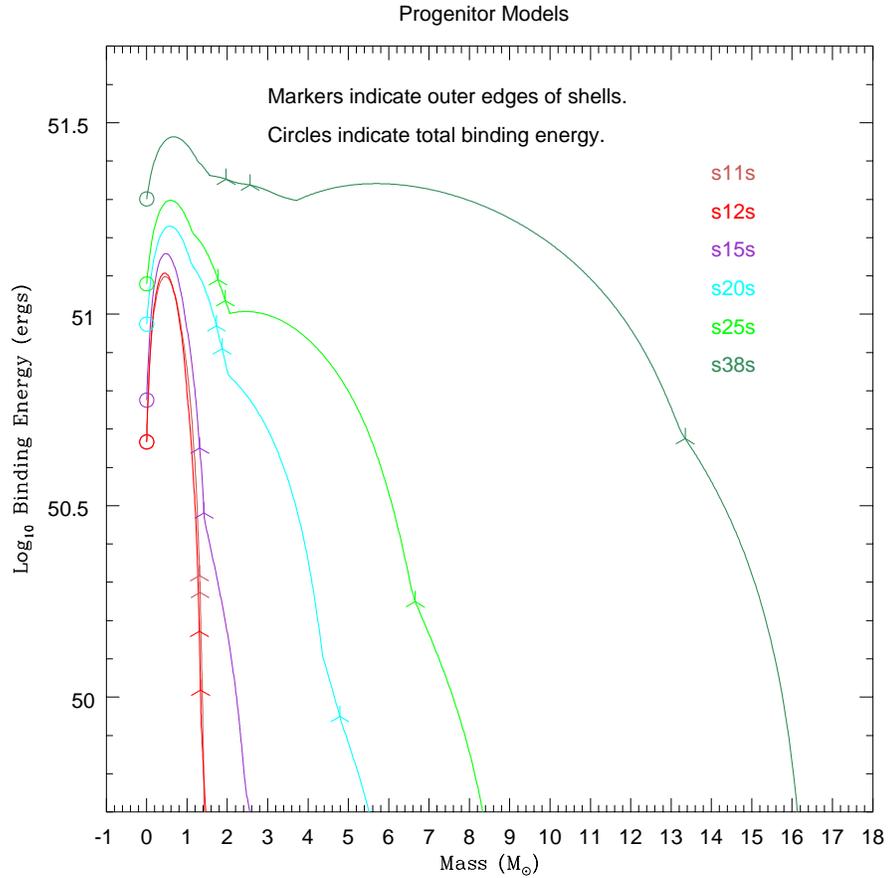}\kern+0in\hfill}

\end{figure}

\section{Hydrodynamic Instabilities and the Supernova Mechanism}

A variety of instabilities and flavors of ``convection'' have been invoked
by theorists over the years to help ignite the supernova explosion. (The study of 
instabilities in the supernova context can be traced to Epstein (1979).)
Mayle \& Wilson (1988) suggested that doubly--diffusive (``neutron finger'') instabilities
in the core enhanced the neutrino luminosity
and turned a dud into an explosion after $\sim$0.5 seconds.
However, Bruenn \& Dineva (1996) show that, because the muon neutrinos decouple from matter just at the time they are
needed to drive the instability, such a salt--finger instability
does not obtain. 
Nevertheless, the Mayle/Wilson luminosity enhancement was only $\sim$20\%,
and this suggested that refinements in the opacity of matter to neutrinos
at high and intermediate densities might be important.  This is still a fruitful lead to follow.

The idea that core overturn driven by negative entropy and lepton gradients persists during
the deleptonization and cooling of the protoneutron star, and that this can boost the driving 
neutrino luminosities, dates back to Burrows, Mazurek, \& Lattimer (1981), but was 
revitalized by Burrows (1987) and Keil, Janka, \& M\"uller (1996).  Whether this is relevant will
be determined only when the best neutrino transport is incorporated in multi--dimensional
protoneutron star cooling calculations.

The neutrino heating rates just behind a stalled shock exceed the cooling rates and
establish the so--called gain region (Bethe \& Wilson 1985; Mayle 1985).  In 1-D, the matter that passes through the shock quickly
leaves this region and may settle onto the core unexceptionally.  However,
in 2--(and presumably 3--)D, the matter can dwell in the gain region for a longer time.  The net result
is a higher entropy and a larger gain region, with more matter less bound behind the shock.
Burrows, Hayes, \& Fryxell (1995) and Herant {\em et al.} (1994) 
describe this situation differently, but each derive that the core
is more unstable to explosion in 2--D than in 1--D, with Herant {\em et al.} (1994)
claiming that overturn is the crucial ingredient.
The equilibrium pre--explosion shock radius is larger in the multi-D
calculations (BHF; Herant \etal\ 1994; Mezzacappa \etal\ 1996; Janka \& M\"uller 1996) 
than in the 1--D calculations, but not all workers agree that
this enables the supernova.  In particular, Mezzacappa {\em et al.} do not obtain an explosion after
$\sim$500 milliseconds and declare their run a dud.  

There are flaws in the calculations of all teams
exploring the role of multi--dimensional effects.  Herant {\em et al.} (1994) use a crude variant of flux--limited diffusion,
cut out the inner 20 kilometers, and use fewer particles in their SPH simulation than may be warranted.
Burrows, Hayes, \& Fryxell (1995) use another  variant of flux--limited diffusion, include the inner core 
(15 kilometers, but follow it in only 1--D), employ PPM and take pains to resolve the flow with many zones 
($500(r) \times 160(\theta)$),
but only in one quadrant, and ignore transport in the angular direction.  Mezzacappa {\em et al.}\,(1996)
couple the transport of their 1--D flux--limited multi--group Lagrangean calculation 
to their 2--D Eulerian PPM calculation (without feedback), 
as a shadow calculation, cut
out the inner 20 kilometers, but move its boundary at a rate determined by the 1--D calculation, start the
2--D calculation a full 106 milliseconds after bounce, and feed the inner 1000 kilometers of the 2--D calculation
according to the 1--D calculation.  Their grid resolution is a marginally adequate $128(r) \times 128(\theta)$, 
but their angular gridding covers
a full 180 degrees.   Janka \& M\"uller (1996) do not follow collapse, but start with a protoneutron star bounded by a stalled shock,
do not follow the hydrodynamics or transport in the inner core, and employ a neutrino light bulb approximation and PPM in the mantle. 
Their focus is on the character of the multi--dimensional flow and the role of neutrino--driven overturn. 
Clearly, though some of the multi--D calculations are more self--consistent than others, all leave something to be desired.
None incorporates general relativity, none adequately handles transport in either the angular or the radial direction, 
none is three--dimensional, and none
is evolved for an adequate duration.  Furthermore, none has correctly treated all the known neutrino processes 
in the core.  In sum, though there has been a great deal of conceptual and numerical progress of late and though there is now
a consensus that neutrinos mediate the energy transfer to an aspherically exploding mantle,
the supernova puzzle in
its particulars remains (Burrows 1996).

\subsection{SN1987A Gamma--Rays and Clumping}

We have recently calculated the emergent gamma--ray and hard X--ray fluxes expected from a
clumpy model of SN1987A (Burrows \& VanRiper 1995).  The purpose of these calculations
was to explain the high flux seen by the Ginga satellite in the 16--28 keV band around
day 614.  Mixed models such as 10HMM (Woosley, Pinto, \& Ensman 1988) and sn14e1 (Nomoto \& Hashimoto 1988),
that could reproduce the early emergence and peaks of the gamma-ray lines and the Compton
continuum, failed to explain the 614--day Ginga data by a factor of from 8 to 25.
Our models employed a variety of artificial radial and angular distributions of $^{56}$Ni, hydrogen, 
and intermediate--mass elements.
We varied the volume filling factors of these components, as well as the density ratios
of the various ``phases.''  Though our prescriptions were ad hoc, 
the models collectively allowed us to explore
the effect of inhomogeneities in the ejecta distributions on the observed hard fluxes.
Our major conclusions were useful: 1) the peaks of the line light curves were fit better if about
half of the $^{56}$Ni produced {\it remained} within the 500--1000 km s$^{-1}$ radius, 2) the filling
factor of hydrogen/helium within 4000  km s$^{-1}$ had to be large, perhaps 50\% -- 60\%, 
and 3) most of 
the $^{56}$Ni and hydrogen/helium could not have been intimately mixed.  Conclusion 1 may have a bearing
on the explosion itself, for it implies that despite the various instabilities, during and
after the explosion, much nickel was {\it not} shot outwards.  
Furthermore, in 1--D, since the explosive nucleosynthesis occurs in a shell around the residue, we would expect that
nickel and heavy elements would be absent from the innermost regions occupied by the protoneutron star.
The wind that is expected (BHF) to emerge from the protoneutron star may also play a role in clearing the deep
interior of the ejecta.
As a consequence, there should be a hole in the line profiles at the lowest radial velocities, though at what
values remains to be calculated (perhaps 500--1000 km s$^{-1}$).  In reality, multi--dimensional instabilities and mixing would
compromise this simple picture, but a hole of some sort should remain.    
This may be consistent with the infrared and optical line profile data of D. Wooden (this volume) 
and R. Stathakis (this volume) and should be a useful constraint on explosion models
and mechanisms.

\section{Neutrino Transport}

Though much of the recent excitement in supernova theory has concerned its
multi--dimensional aspects, neutrino heating and transport are still central
to the mechanism.  The coupling between matter and radiation in the semi--transparent
region between the stalled shock and the neutrinospheres determines the viability
and characteristics of the explosion.   Unfortunately, this is the most problematic regime.
Diffusion algorithms and/or flux--limiters do not adequately reproduce the effects of variations
in the Eddington factors and the spectrum as the neutrinos decouple.  Hence, a multi--group full transport
scheme is desirable.  

To address the issues surrounding neutrino transport, we have recently
created a neutrino transport code using the program {\bf Eddington} developed
by Eastman \& Pinto (1993).  This code solves the full
transport equation using the Feautrier approach,
is multi--group, is good to order v/c, and does not employ
flux limiters.  The $\nu_e$s, $\bar{\nu}_{e}$s, and ``$\nu_\mu$s''
are handled separately
and coupling to matter is facilitated with accelerated lambda iteration (ALI).
By default, we employ 40 energy groups from 1 MeV to either 100 MeV ($\bar{\nu}_{e}$ and ``$\nu_\mu$s'')
or 230 MeV ($\nu_e$) and from a few to 200 angular groups, depending on the number of tangent rays at the given radial zone.
In this way, the neutrino angular distribution function
and all the relevant angular moments (0'th through 3'rd) are calculated to high precision, for every energy group.

The effect of the full Feautrier
scheme vis--a--vis previous calculations 
(BHF; Mezzacappa \& Bruenn 1993a,1993b; Mezzacappa \etal\ 1996; Wilson 1985; Myra \& Burrows 1990)
is being benchmarked and calibrated.
However, we have already obtained several interesting results.
Since the annihilation of $\nu$--$\bar{\nu}$ pairs into $e^+$--$e^-$ pairs depends upon the 0'th, 1'st, and 2'nd
angular moments of the neutrino angular distribution function, as well as upon the neutrino spectra, we can and have
calculated the rate of energy deposition via this process {\it exactly}, though in the context of previous model runs (BHF) 
and ignoring general relativity.
Figure 3 compares the energy deposition rates due to neutrino pair annihilation with those of the charged--current absorption processes. 
 
\begin{figure}
\vspace{3.5in}
 
\caption{The energy deposition rates (in ergs gm$^{-1}$ s$^{-1}$) versus radius (in centimeters) 30 milliseconds after bounce, due to
neutrino pair annihilation (dashed) and charged--current absorption (top two solid curves).  The $\bar{\nu}_e$--$\nu_e$
and $\bar{\nu}_\mu$--$\nu_\mu$ annihilation processes are distinguished (the latter is above the former).  
Heating by $\nu_\mu$s due to other than $\bar{\nu}_\mu$--$\nu_\mu$ annihilation is also shown (lowest solid line).
The full transport machinery described in the text was used
to obtain the numbers plotted.
}

\label{fig-3}
\hbox to\hsize{\hfill\includegraphics{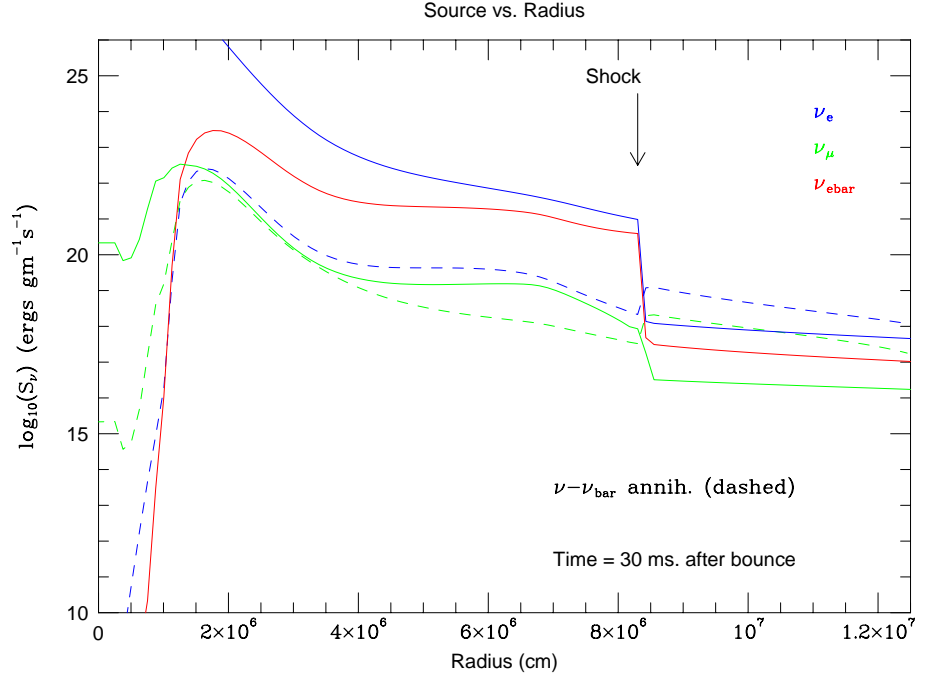}\kern+0in\hfill}

\end{figure}
 
The $\nu_e + \bar{\nu}_e \rightarrow  e^+ + e^-$ and $\nu_\mu + \bar{\nu}_\mu \rightarrow  e^+ + e^-$
energy deposition rates in the shocked region are no more than 0.01 and 0.001, respectively, those of the dominant charged--current
processes,  $\nu_e + n \rightarrow  e^- + p$ and $\bar{\nu}_e + p \rightarrow  e^+ + n$.  
However, in the 
unshocked region ahead of the shock, depending upon the poorly--known $\nu$--nucleus
absorption rates, the $\nu$--$\bar{\nu}$ annihilation rate can be competitive, though
it is still irrelevant to the supernova.

It is thought that neutrino--electron scattering and inverse
pair annihilation are the processes most responsible for the energy
equilibration of the $\nu_\mu$'s and their emergent spectra.  However,
recent calculations imply that the inverse of nucleon--nucleon
bremsstrahlung ({\it e.g.}, $n + n \rightarrow n + n + \nu \bar{\nu}$) is
also important in equilibrating the $\nu_\mu$'s (Suzuki 1993).  This
process has not heretofore been incorporated in supernova simulations.
Our preliminary estimates suggest that inverse bremsstrahlung
softens the emergent $\nu_\mu$ spectrum, since the bremsstrahlung 
source spectrum is softer than that of pair annihilation.
In addition, given the large $\nu_\mu$ scattering albedo, one must 
properly distinguish absorption from scattering, in ways not possible 
with a flux--limiter.  Since 
the relevant inelastic neutral--current processes are stiff functions of neutrino
energy, these transport issues bear directly upon the viability of neutrino 
nucleosynthesis ({\it e.g.}, of \nuc{11}{B} and \nuc{19}{F}) (Haxton 1988; Woosley \etal\ 1990). 

The new code allows us to calculate the difference between the flux spectrum ($h_{\nu}$) and the
energy density spectrum ($j_{\nu}$).  The latter couples to matter and drives the supernova in the
neutrino mechanism,  while the
former, or some variant of it, is frequently substituted for the latter in diffusion
codes.  Since matter--neutrino cross sections are higher for higher--energy
neutrinos, the energy density spectrum is always harder than the flux spectrum.
This hardness boosts the neutrino heating rates in the semi--transparent
region.   To illustrate this effect, in Figure 4 the ratio $j_{\nu_e}$/$h_{\nu_e}$ is plotted versus neutrino energy at
a time 30 milliseconds after bounce.   The shock is then at 124 kilometers.  It is clear that
the ratio effect can be interesting.  However, it is most pronounced in the cooling region below
the gain region and tapers off as the shock is approached.  Mezzacappa {\em et al.} (1996),
in particular, have highlighted this correction, but self--consistent calculations from
collapse to explosion, using the Feautrier or Boltzmann techniques (in
principle equivalent), are needed, given the notorious feedbacks in the supernova problem.  
The same effect may be important in driving the protoneutron star wind (BHF) 
thought to be the site of the r--process (Woosley \& Hoffman 1992; Qian \& Woosley 1996).  Qian \& Woosley (1996) suggest
that an extra heating source in the wind acceleration region may help
establish the conditions for a successful r--process.  Such a ``source'' may be a natural consequence of the proper
handling of neutrino transport above the neutrinosphere.
Indeed, full transport calculations of r--process winds and the supernova, even in 1--D, 
will be illuminating.

\begin{figure}
\vspace{3.5in}

\caption{$j_{\nu}/h_{\nu}$ versus $\epsilon_\nu$ for electron neutrinos 30 milliseconds after the bounce
of a 15 M$_\odot$ core, using the code of Burrows \& Pinto (1997). }  

\label{fig-4}
\hbox to\hsize{\hfill\includegraphics{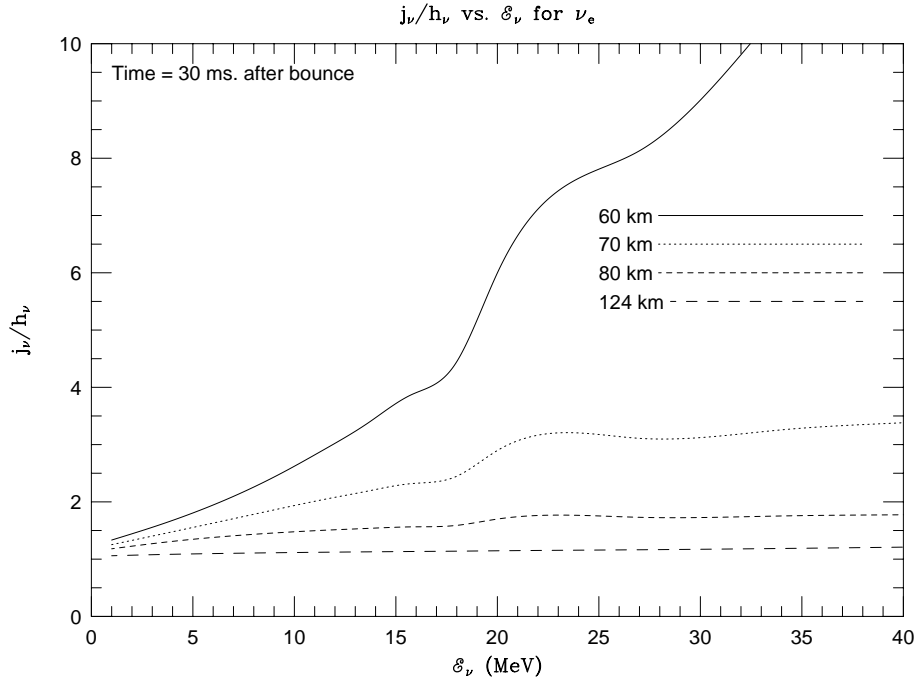}\kern+0in\hfill}

\end{figure}

\section{Pulsar Proper Motions, Natal Kicks, and Gravitational Radiation}

A new pulsar distance scale (Taylor \& Cordes 1993), recent pulsar proper
motion data (Harrison, Lyne, \& Anderson 1993),
and the recognition that pulsar
surveys are biased towards low speeds (Lyne \& Lorimer 1994; Fryer, Burrows, \& Benz 1997) imply that radio
pulsars are a high-speed population.
Mean three-dimensional galactic speeds of 450$\pm$90 km s$^{-1}$
have been estimated (Lyne \& Lorimer 1994), with measured transverse speeds of
individual pulsars ranging from
zero to $\sim$1500 km s$^{-1}$.   Impulsive mass loss in
a spherical supernova explosion that occurs in a binary can impart
to the nascent
neutron star a substantial kick that reflects its progenitor's
orbital speed (Gott, Gunn, \& Ostriker 1970).
However, theoretical studies of binary evolution through the
supernova phase have difficulty reproducing
velocity distributions with the required mean and dispersion (Fryer, Burrows, \& Benz 1997).
This implies that neutron stars receive an extra kick at birth.

In the past, an off-center (and rapidly rotating) magnetic
dipole (Harrison \& Tademaru 1975)
and anisotropic neutrino radiation
(Woosley 1987; Chugai 1984) have been
invoked to accelerate neutron stars.
A 1\% net asymmetry in the neutrino radiation of a neutron
star's binding energy
results in a $\sim$300 km s$^{-1}$ kick.
However, Burrows \& Hayes (1996) have recently demonstrated
that if the
collapsing Chandrasekhar core is mildly asymmetrical,
the young neutron star can receive a large impulse during the
explosion in which
it is born.
In those calculations, rocket-like mass motions, not neutrinos,
dominated the recoil, which reached $\sim$530 km s$^{-1}$.
Such a speed is large,
but is only $\sim$2\% of that
of the supernova ejecta.
This asymmetry/recoil
correlation seems generic.
However, whether such asymmetries are themselves generic has yet
to be demonstrated.
Recent calculations of convection
during shell oxygen and silicon burning (Bazan \& Arnett 1994)
suggest that the initial density, velocity, and
composition asymmetries
might indeed be interesting.

The impulse delivered to the core depends upon the dipole moments
of the angular distribution of both the
envelope momentum and the neutrino luminosity.  The gravitational
waveform depends upon the corresponding
quadrupole moments.  Curiously, using the standard quadrupole
formula, Burrows \& Hayes (1996) derived that due to
the intense and anisotropic early neutrino burst,
the neutrino contribution to the metric strain, $h^{TT}_{zz}$,
can dominate
during the early post-bounce epoch.
This is true despite the fact
that the neutrinos do not dominate the recoil and is a consequence
of their relativistic nature.
Figure 5 depicts the power spectrum of gravitational radiation emitted in two of the theoretical models
of anisotropic collapse in Burrows \& Hayes (1996).  The only difference between the models was in the character of the 
initial pertrubations to the collapsing Chandrasekhar core.

\begin{figure}
\vspace{4.5in}
 
\caption{The power spectrum (in strain Hz$^{-1/2}$) versus frequency (in Hz) 
at a distance of 10 kiloparsecs for two of the models
in the study of Burrows \& Hayes (1996).  Superposed is the putative sensitivity curve of the 2'nd--generation
LIGO detector. 
}
 
\label{fig-5}
\hbox to\hsize{\hfill\includegraphics{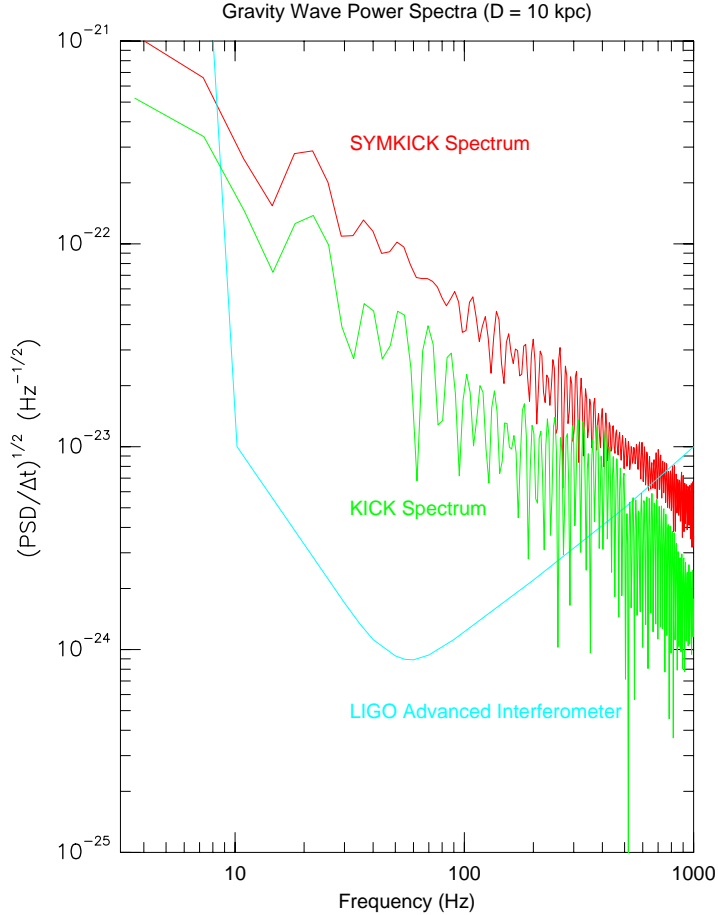}\kern+4in\hfill}
 
\end{figure}

The gravitational waves are radiated between 10 and 500 Hz
and h$^{TT}_{zz}$ does
not go to zero with time.  Hence, there may be ``memory''
(Braginsky \& Thorne 1987) in the
gravitational waveform from a protoneutron star that is
correlated with
its recoil and neutrino emissions.
This memory is a distinctive
characteristic of asymmetric collapse and explosion.

\section{Conclusions}

We have found and documented a new richness in the theory of 
supernova explosions that stems from their multi--dimensional
character, but have yet to achieve the type of quantitative
successes that are the theory's modern goals.  A reappraisal of the neutrino
sector reveals that new processes and algorithms may alter our view 
of neutrino--driven explosions and of the spectra of the emergent neutrinos.
Hydrodynamic instabilities in the core and mantle of the collapsed Chandrasekhar mass
suggest that the protoneutron star is given kicks and torques (spins) at birth,
that the supernova gravitational radiation signature has significant power at low frequencies,
and that the debris field is grossly aspherical.  Whatever emerges from these new 
analyses, one thing is clear: Nature is stingy with its secrets that we should still be so unenlightened
on the 40'th anniversary of Burbidge, Burbidge, Fowler, \& Hoyle (1957)
and the 10'th anniversary of SN1987A.

\acknowledgements 
 
I would like to thank Willy Benz, Chris Fryer, Bruce Fryxell, 
John Hayes, Rob Hoffman, Evonne Marietta, Tony Mezzacappa, Phil Pinto,    
Raylee Stathakis, Diane Wooden, and Stan Woosley for useful conversations and insights.
This work was supported under NSF grant AST96-17494.

\end{document}